\def\ROSAT{{\it ROSAT\/\ }}

\def\ASCA{{\it ASCA\/\ }}

\def\SAX{{\it SAX\/\ }}

\def\Einstein{{\it Einstein\/\ }}

\def\ltsima{$\; \buildrel < \over \sim \;$}
\def\simlt{\lower.5ex\hbox{\ltsima}}
\def\gtsima{$\; \buildrel > \over \sim \;$}
\def\simgt{\lower.5ex\hbox{\gtsima}}

\documentstyle[twoside,fleqn,psfig]{an98-art}

\input psfig.sty

\begin{document}

\begin{titlepage}
\setcounter{page}{1}
\makeheadline

\title {Ultrasoft Narrow-Line Seyfert~1 Galaxies and X-ray Surveys}

\author{{\sc Niel Brandt}, State College, Pennsylvania, USA\\
\medskip
{\small The Pennsylvania State University} \\
\bigskip
{\sc Thomas Boller}, Garching, Germany\\
\medskip
{\small Max-Planck-Institut f\"ur Extraterrestrische Physik} \\
}

\date{Received; accepted } 
\maketitle

% ---------------------------------------------

\summary
We describe how recent X-ray surveys have led to advances in the 
understanding of ultrasoft narrow-line Seyfert~1 galaxies.
The number of known ultrasoft narrow-line
Seyfert~1s has increased greatly in recent years due to X-ray
surveys, and it is now possible to obtain 
high quality 0.1--10~keV spectral and
variability measurements for a large number of 
these galaxies. 
We generalize some of the correlations between X-ray
properties and optical emission line 
properties, focusing on how the \ROSAT band spectral slope
appears to be directly connected to the Boroson \& Green (1992) 
primary eigenvector. 
We discuss how ultrasoft narrow-line Seyfert~1s may well have
extremal values of a primary physical parameter, and we
describe new projects that should further 
improve our understanding of these 
extreme representatives of Seyfert activity. 
END

% ---------------------------------------------

\keyw
galaxies: Seyfert -- galaxies: active -- X-rays: galaxies. 
END

% ---------------------------------------------

\AAAcla
Narrow-line Seyfert~1 galaxies (NLS1)
END
\end{titlepage}

% ---------------------------------------------

\kap{Ultrasoft Narrow-Line Seyfert~1s: New Objects and Quality X-ray Measurements }

\noindent 
While Seyfert~1 type galaxies with extremely strong soft X-ray excess 
components (relative to their hard X-ray power-law components)
had been identified as a class
based on data from \Einstein and earlier X-ray satellites
(e.g. C\' ordova et~al. 1992; Puchnarewicz et~al. 1992), observations
with {\it ROSAT}, \ASCA and \SAX have revolutionized the study of 
these galaxies. First of all, \ROSAT has vastly 
{\bf increased the number} of ultrasoft 
Seyferts known. At least 50 new ultrasoft
Seyferts have been found using \ROSAT data (one of the largest 
catalogs of new ultrasoft Seyferts is presented by Grupe 1996),
and more are being discovered each month. Many of these are 
bright and suitable for detailed follow-up studies at X-ray
and other wavelengths. The best available estimates
suggest that there are $\simgt 1000$ ultrasoft Seyferts in the 
\ROSAT all-sky survey (RASS) above a flux level of 
$5\times 10^{-13}$ erg cm$^{-2}$ s$^{-1}$, so new ultrasoft 
Seyfert identifications should continue for some time. 

% ----------------------------------------------------------------

\begin{figure}
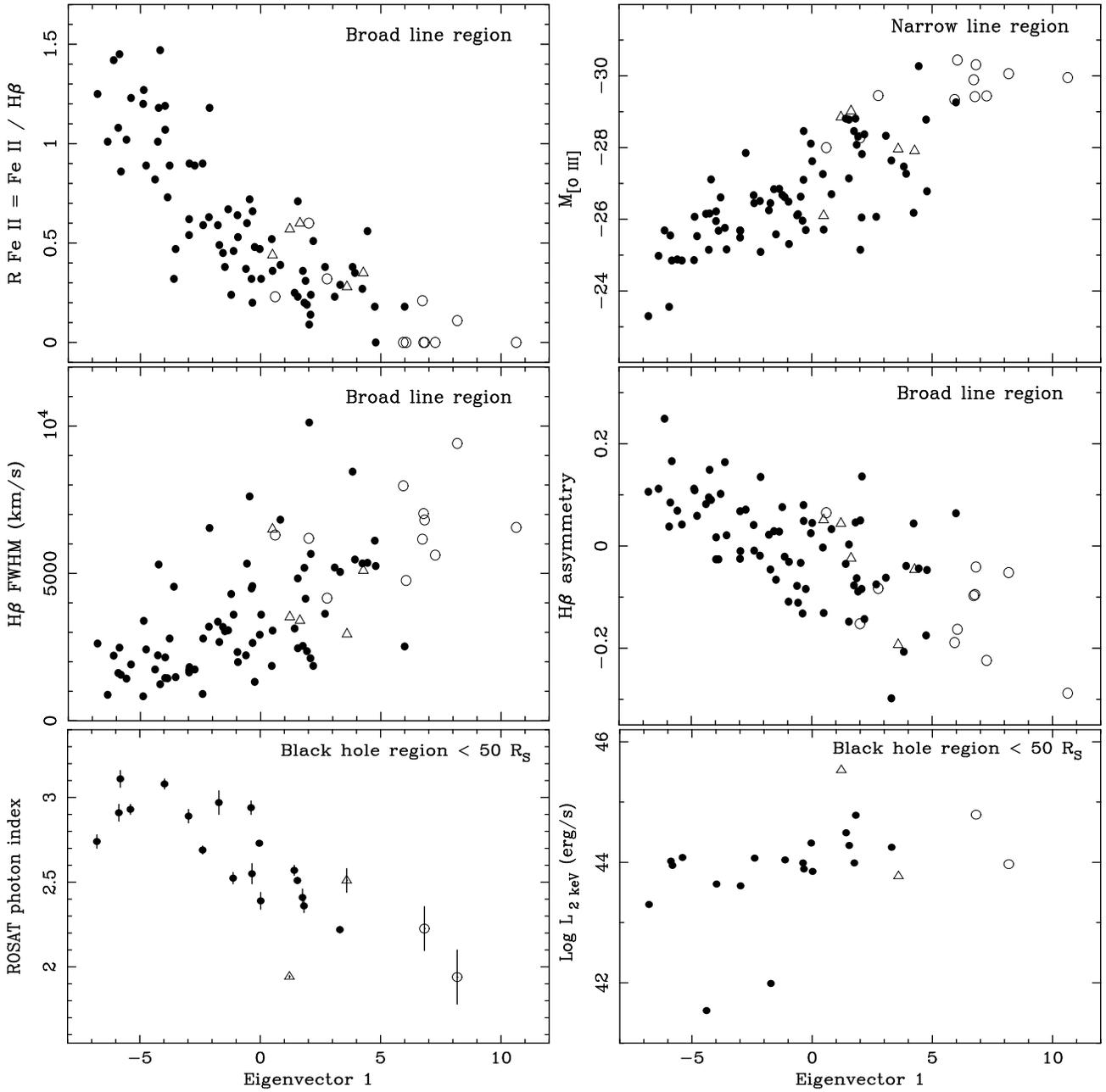

\hbox{
\psfig{figure=rfeii.ps,width=0.49\textwidth,height=0.23\textheight,angle=270}
\psfig{figure=oiii-lum.ps,width=0.49\textwidth,height=0.23\textheight,angle=270}
}
\hbox{
\psfig{figure=hb-fwhm.ps,width=0.49\textwidth,height=0.23\textheight,angle=270}
\psfig{figure=hb-asym.ps,width=0.49\textwidth,height=0.23\textheight,angle=270}
}
\hbox{
\psfig{figure=bgxray-slope.ps,width=0.49\textwidth,height=0.23\textheight,angle=270}
\psfig{figure=bgxray-lum-2keV.ps,width=0.49\textwidth,height=0.23\textheight,angle=270}
}
\caption{QSO optical emission line and X-ray properties versus 
the value of the BG92a primary eigenvector. The QSO
in these plots are those from BG92a, and the
X-ray measurements have been taken from L97. This
plot is similar to that presented in BG92b,
although the direct correlations of the X-ray data with the 
primary eigenvector are shown here for the first time. 
Solid dots are radio-quiet QSO, open triangles are
core-dominated radio-loud QSO, and open circles are 
lobe-dominated radio-loud QSO. In the upper right corner
of each panel we state the region of the QSO being probed by the
quantity along the corresponding ordinate, and we note that size
scales ranging from hundreds of parsecs to tens of Schwarzschild
radii have correlated properties. In particular, it is clear that the
property driving the eigenvector can strongly affect the
appearance of the luminous and 
rapidly variable X-ray emission from the 
black hole region (within $\sim 50$ Schwarzschild radii of 
the black hole itself).
In the lower left panel we have not plotted the QSO
PG~1001+054 because its \ROSAT photon index has a very large
error bar (although the position of this QSO does agree well
with the pictured trend). Also in this panel we note that the
outlying open triangle is 3C273, which has a time-variable 
photon index at \ROSAT energies. This may well be due to
time-variable ionized absorption (see Guainazzi \& Piro 1997), 
and 3C273 also has some blazar-like characteristics. 
In the lower right panel the two outlying solid dots
are `X-ray weak QSO' (see L97 for discussion). 
Following section~4 of Kellermann et~al. (1994), we take 
PG~1211+143 to be radio-quiet. 
}
\end{figure}

% ----------------------------------------------------------------

{\it ROSAT}, \ASCA and \SAX observations have also greatly
{\bf improved the quality} of the X-ray data
available for ultrasoft Seyferts. Precise measurements of their 
0.1--10~keV spectral and variability properties are now 
possible. Comparisons of these precise X-ray measurements with
data at other wavelengths has lead to the discovery of
strikingly clear relations between the X-ray properties of
Seyferts and QSO and their optical emission line
properties (e.g. Boller, Brandt \& Fink 1996, hereafter BBF96; 
Laor et~al. 1997, hereafter L97). Most notably, 
the FWHM of the optical H$\beta$ line appears
to be anticorrelated with the slope of the \ROSAT band X-ray
continuum (which serves as a measure of the 
strength of the soft X-ray excess compared to the power law). 
{\it All\/} of the softest Seyfert~1s known 
appear to be `narrow-line' Seyfert~1s 
(hereafter NLS1; e.g. Osterbrock \& Pogge 1985) 
with typical H$\beta$ FWHM in the range 
500--2000~km~s$^{-1}$. Ultrasoft NLS1 lie at one extreme of the
aforementioned anticorrelation, and they are sometimes called 
`I~Zwicky~1' type objects after one of the most famous NLS1. 
Other relations between X-ray properties and optical 
emission line properties have also been found 
(see BBF96, L97 and references therein), and we will 
describe these within a generalized framework in 
the next section. 

% ---------------------------------------------

\kap{Generalizing the Correlations Between X-ray Emission and Optical Emission Lines }

\noindent
Boroson \& Green (1992a, hereafter BG92a) have 
used a principal component analysis
(PCA) to identify a fascinating cluster of QSO optical emission 
line properties that vary together in a highly coordinated manner,
the `primary eigenvector' of their PCA. The 
relevant properties include
Fe~{\sc ii} strength, 
[O~{\sc iii}] strength, 
H$\beta$ FWHM
and H$\beta$ asymmetry 
[see the top four panels of Fig.~1;
for the H$\beta$ asymmetry panel, 
H$\beta$ lines that have excess light in their blue (red) 
wings are positive (negative)].
BG92a and Boroson (1992) use the [O~{\sc iii}] emission to 
argue that the property driving the eigenvector cannot be 
orientation. They also note that radio-loud QSO tend to lie
toward the positive extreme of the eigenvector and suggest that the
property driving the eigenvector is linked 
to the radio `volume control' of QSO. NLS1 lie toward the
negative extreme of the eigenvector.   

In the lower two panels of Fig.~1, we show correlations of \ROSAT
power-law photon index and 2~keV luminosity 
versus the BG92a primary eigenvector. A highly
significant anticorrelation between \ROSAT photon index and the 
eigenvector is apparent in the lower left panel. A 
Spearman rank-order correlation
gives a Spearman $r_s$ value of $-0.834$, which is significant 
with over 99.9 per cent confidence (we use all 23 of the L97 QSO
in this calculation). This anticorrelation is {\it stronger\/} than
any of the L97 correlations/anticorrelations of X-ray 
properties with emission line widths and ratios, suggesting that 
it is more fundamental and supporting the idea that the 
eigenvector has an important physical meaning.
We note that the \ROSAT photon index was not used in the PCA 
calculation of the primary eigenvector, so the abscissa and
ordinate in the lower left panel are independent quantities
(this is not the case for the upper four panels but BG92a
show that the correlations are not artificially induced in
this manner).  
The energetically important and rapidly 
variable X-ray emission is thought to
be formed within $\sim 50$ Schwarzschild radii ($R_{\rm S}$) of 
the supermassive black hole itself. {\it Thus it appears that the property
which ultimately drives the eigenvector also originates within\/}
$\sim 50 R_{\rm S}$ (or at least can strongly influence the 
appearance of the X-ray emission from this region). 
We comment that PG QSO were selected by their UV excesses
and most are relatively `clean' systems in that they do not 
appear to have large amounts of 
intrinsic obscuration along the line of sight
to the active core (although there are a few exceptions to this
general rule). It appears unlikely that complex X-ray 
absorption effects are confusing the 
measurements of the X-ray continuum shape
(see L97 for details of the \ROSAT fits).

Corbin (1993) correlated the 2~keV X-ray luminosities of
55 PG QSO with their optical emission line properties from
BG92a, and he argued that the 2~keV luminosity was linked to
the BG92a primary eigenvector. We also see such a correlation
in the lower right panel of Fig.~1 (it has a Spearman 
$r_{\rm s}=0.468$ corresponding
to a correlation probability of 97.6 per cent). However,
this correlation is significantly weaker than the one in the
lower left panel, suggesting that the low 2~keV luminosities
at the negative end of the eigenvector
are only a secondary consequence of the steep X-ray slope
(in agreement with L97; this remains true even if we
exclude the two `X-ray weak QSO'). The 0.3~keV luminosity
does not show a significant correlation with the BG92a 
primary eigenvector, consistent with the idea
that the X-ray spectral shape, rather than
the X-ray luminosity, is the primary quantity 
responsible for the X-ray correlations. 

Inspection of Fig.~1 also makes it clear why soft X-ray surveys
are very effective at finding new 
ultrasoft NLS1: soft X-ray selection will
preferentially choose the steep X-ray spectrum 
Seyferts and QSO with negative 
values of the BG92a primary eigenvector. 
These Seyferts and QSO tend to have
narrow H$\beta$ FWHM 
as well as
strong Fe~{\sc ii}, 
weak [O~{\sc iii}] and
blue asymmetric H$\beta$ profiles. 

% ---------------------------------------------

\kap{Determining the Physical Parameter that Drives the Correlations}
 
\noindent
Ultrasoft NLS1 have extreme optical emission line properties and 
extreme X-ray properties. The fact that they persistently 
tend to lie at the ends of distributions 
of Seyfert quantities suggests that they
may have extremal values of 
some important, underlying Seyfert physical
parameter. BG92a argued that the parameter 
driving the eigenvector is not orientation
but rather an intrinsic property. We have argued above 
that the parameter originates close to the supermassive black 
hole and can induce energetically important changes in the
X-ray spectral energy distribution (these X-ray clues help to
restrict the number of possibilities that must be considered
for the driving parameter). It is unlikely that black hole mass
is the sole driver of the correlations due to the fact that NLS1
characteristics are observed in sources spanning an 
extremely wide range of luminosity. However, the  
fraction of the Eddington rate at which the supermassive 
black hole is accreting (see BG92b)
or
the spin rate of the black hole 
might plausibly drive the correlations. 
If we can clearly determine the driving parameter this would be
an important advance, since we would then be able to 
study the observational consequences of a difference
in a primary driver of these accretion powered sources. 
We stress that the correlations discussed above are the 
{\it strongest\/} ones that emerge when large samples of 
Seyferts and QSO are systematically studied (that is, we
are not trying to explain mere `2 per cent effects'). 

How might one clearly determine the driving parameter? One possible
avenue for progress is to carefully and systematically search for 
other properties that correlate with the BG92a eigenvector. 
Hard X-rays offer several properties that can be tested for
correlation. The intrinsic hard X-ray continuum slope is one of the
most basic and important of these, since it is thought to probe
the temperature and Thomson depth of the accretion disk corona. 
The currently available data suggest
that there is a correlation with NLS1 having steeper
intrinsic continuum slopes on average
(Brandt, Mathur \& Elvis 1997; although the correlation
appears weaker than that in soft X-rays).
Hard X-ray spectral slope measurements for NLS1 with
strong optical Fe~{\sc ii} emission will also 
help to constrain models of Fe~{\sc ii} line
formation (e.g. sect.~3.2 of Brandt et~al. 1997
and references therein). Iron~K line 
and Compton reflection continuum
properties could also be checked for correlations.
Some interesting iron~K line results for ultrasoft NLS1 have
already emerged (e.g. Comastri et~al. 1997), but better data
are needed before general conclusions can be drawn. 
X-ray variability, while often difficult to
rigorously quantify, offers another promising set of
properties that can be tested for correlation with the
BG92a eigenvector. Several NLS1 have certainly shown
dramatically rapid, large-amplitude and nonlinear 
X-ray variability (see Boller et~al. 1997 for one of
the most interesting examples and references; also
see Fiore 1997).

% , and more systematic comparisons of 
% variability properties are underway. 

Finally, it is important to check for loopholes in
some of the argumentation 
that is often taken for granted. For
example, [O~{\sc iii}] emission has been
used to argue against an orientation interpretation
of the BG92a eigenvector (see above). However, the
relevant arguments rely on the assumption that 
[O~{\sc iii}] is an isotropic property. This common
assumption has been called into question
(e.g. Hes, Barthel \& Fosbury 1993; Baker 1997). Measurements
of [O~{\sc ii}] emission, which has a lower ionization
potential and critical density than [O~{\sc iii}],
would be useful for critically examining the 
orientation interpretation. 

% Radio loudness connection --- need more work here

% ---------------------------------------------

\acknowledgements
WNB gratefully acknowledges the support of the Pennsylvania State
University and the Harvard-Smithsonian Center for Astrophysics.
We thank T. Boroson for providing the eigenvector values for the
individual QSO, and we thank M. Elvis, A. Laor, B. Wilkes and the 
members of the Harvard QED team for helpful discussions. 

%
%%%%% References
\refer
\aba
\rf{Baker J.C., 1997, MNRAS, 286, 23}
\rf{Boller Th., Brandt W.N., Fink H., 1996, A\&A, 305, 53 (BBF96)}
\rf{Boller Th., Brandt W.N., Fabian A.C., Fink H., 1997, MNRAS, 289, 393}
\rf{Boroson T.A., 1992, ApJ, 399, L15}
\rf{Boroson T.A., Green R.F., 1992a, ApJS, 80, 109 (BG92a)}
\rf{Boroson T.A., Green R.F., 1992b, 
in Holt S.S., Neff S.G., Urry C.M., eds, 
Testing the AGN Paradigm.
AIP Press, New York, p. 584 (BG92b)}
\rf{Brandt W.N., Mathur S., Elvis M., 1997, MNRAS, 285, L25}
\rf{Brandt W.N., Mathur S., Reynolds C.S., Elvis M., 1997, 
MNRAS, in press (astro-ph/9707216)}
\rf{Comastri A., et~al., 1997, A\&A, submitted}
\rf{Corbin M.R., 1993, ApJ, 403, L9}
\rf{C\' ordova F.A., Kartje J.F., Thompson R.J., Mason K.O., 
Puchnarewicz E.M., Harnden F.R., 1992, ApJS, 81, 661}
\rf{Fiore F., 1997, Mem. Soc. Ital. Astr. Soc., 68, 119}
%
% \rf{Goodrich R.W., 1989, ApJ, 342, 224}
%
\rf{Grupe D., 1996, PhD thesis, Uni-Sternwarte G\"ottingen}
\rf{Guainazzi M., Piro L., 1997, Astron. Nachr., in press (astro-ph/9708260)}
\rf{Hes R., Barthel P.D., Fosbury R.A.E., 1993, Nature, 362, 326}
\rf{Kellermann K.I., Sramek R., Schmidt M., Green R.F.,
Shaffer D.B., 1994, AJ, 108, 1163}
\rf{Laor A., Fiore F., Elvis M., Wilkes B.J.,
McDowell J.C., 1997, ApJ, 477, 93 (L97)}
\rf{Osterbrock D.E., Pogge R.W., 1985, ApJ, 297, 166}
\rf{Puchnarewicz E.M., Mason K.O., C\' ordova F.A., 
Kartje J., Branduardi-Raymont G., Mittaz J.P.D.,
Murdin P.G., Allington-Smith J., 1992, MNRAS, 256, 589}
\abe
%%%%% End of references

%
%%%%% Address of the authors
\addresses
\rf{Niel Brandt, 
Department of Astronomy and Astrophysics, 
The Pennsylvania State University, 
525 Davey Lab, 
University Park, PA 16802,
USA,
e-mail: niel@astro.psu.edu}
\rf{Thomas Boller, Max-Planck-Institut f\"ur Extraterrestrische
Physik, 85740 Garching, Germany, 
e-mail: bol@rosat.mpe-garching.mpg.de}
END
%
%%%%% End of address

\end{document}